\documentclass[preprint,preprintnumbers,showpacs,amsmath,amssymb]{revtex4-1}
\usepackage{graphicx}
\usepackage{epsfig}
\usepackage{amssymb} 
\usepackage{graphicx}
\usepackage{dcolumn}
\usepackage{bm}
\newcommand{\ba}{\begin{array}}
\newcommand{\ea}{\end{array}}
\def\br{\begin{eqnarray}}
\def\er{\end{eqnarray}}
\def\be{\begin{equation}}
\def\ee{\end{equation}}

\def\({\left(}
\def\){\right)}

\begin{document}


\title{Non-perturbative QCD effects in forward scattering at LHC}

\author{C.A.S. Bahia, M. Broilo, and E.G.S. Luna}
\affiliation{Instituto de F\'isica, Universidade Federal do Rio Grande do Sul, Caixa Postal 15051, 91501-970, Porto
Alegre, RS, Brazil }  
 

\begin{abstract}
We study infrared contributions to semihard parton-parton interactions by considering an effective charge whose finite
infrared behavior is constrained by a dynamical mass scale. Using an eikonal QCD-based model in order to connect this
semihard parton-level dynamics to the hadron-hadron scattering, we obtain predictions for the proton-proton ($pp$) and
antiproton-proton ($\bar{p}p$) total cross sections, $\sigma_{tot}^{pp,\bar{p}p}$, and the ratios of the real to imaginary
part of the forward scattering amplitude, $\rho^{pp,\bar{p}p}$. We discuss the theoretical
aspects of this formalism and consider the phenomenological implications of a class of energy-dependent form factors
in the high-energy behavior of the forward amplitude. We introduce integral dispersion relations specially tailored to
relate the real and imaginary parts of eikonals with energy-dependent form factors. Our results, obtained using a group
of updated sets of parton distribution functions (PDFs), are consistent with the recent data from the TOTEM, AUGER and
Telescope Array experiments.
\end{abstract}


\maketitle

\section{Introduction}

The study of hadron-hadron total cross sections has been a subject of intense theoretical and experimental interest.
The recent measurements of $pp$ elastic, inelastic and total cross sections at the LHC by the TOTEM Collaboration
\cite{TOTEM001,TOTEM002,TOTEM003,TOTEM004} have enhanced the interest in the subject and become a pivotal source of
information for selecting models and theoretical methods. At present one of the main theoretical approaches for the
description of the observed increase of hadron-hadron total cross sections is the QCD-inspired formalism
\cite{durand,luna001,luna009,giulia001}. In this approach the energy dependence of the total cross section
$\sigma_{tot}(s)$ is obtained from the QCD using an eikonal formulation compatible with analyticity and
unitarity constraints. More precisely, the behavior of the forward observables $\sigma_{tot}(s)$ and $\rho(s)$ is
derived from the QCD parton model using standard QCD cross sections for elementary parton-parton processes, updated
sets of quark and gluon distribution functions and physically-motivated cutoffs which restrict the parton-level
processes to semihard ones. These semihard processes arise from hard scatterings of partons carrying very small
fractions of the momenta of their parent hadrons, leading to the appearance of jets with transverse energy $E_{T}$
much smaller than the total energy $\sqrt{s}$ available in the hadronic collision. In this picture the scattering of
hadrons is an incoherent summation over all possible constituent scattering and the increase of the total cross
sections is directly associated with parton-parton semihard scatterings. The high-energy dependence of the cross
sections is driven mainly by processes involving the gluon contribution, since it gives the dominant contribution at
small-$x$.

However, despite this scenario being quantitatively understood in the framework of perturbative QCD, the
non-perturbative character of the theory is also manifest at the elementary level since at high energies the soft and
the semihard components of the scattering amplitude are closely related \cite{gribov001}. Thus, in considering the
forward scattering amplitude,
it becomes important to distinguish between semihard gluons, which participate in hard parton-parton scattering, and
soft gluons, emitted in any given parton-parton QCD radiation process.

Fortunately, our task of describing forward observables in hadron-hadron collisions, brin\-ging up information about
the infrared properties of QCD, can be properly addressed by considering the possibility that the non-perturbative
dynamics of QCD generate an effective gluon mass. This dynamical gluon mass is intrinsically related to an infrared
finite strong coupling constant, and its existence is strongly supported by recent QCD lattice si\-mu\-la\-ti\-ons
\cite{lqcd} as well as by phenomenological results \cite{luna001,luna009,durao}.
More specifically, a global description of $\sigma_{tot}^{pp,\bar{p}p}(s)$ and $\rho^{pp,\bar{p}p}(s)$ can succeed in a
consistent way by introducing a non-perturbative QCD effective charge in the calculation of the parton-level processes
involving gluons, which dominates at high energy and determines the asymptotic behavior of hadron-hadron cross
sections. 

With this background in mind, the main purpose of this paper is to explore the non-perturbative dynamics of QCD in
order to describe the total cross section, $\sigma_{tot}(s)$, and the ratio of the real to imaginary part of the
forward scattering amplitude, $\rho(s)$, in both $pp$ and $\bar{p}p$ channels, assuming the eikonal representation
and the unitarity condition of the scattering matrix. In our analysis we introduce a new class of energy-dependent
form factors which represent the overlap density for the partons at impact parameter $b$. We relate the real and
imaginary parts of the eikonal by means of suitable dispersion relations for amplitudes with energy-dependent form
factors. We also explore the effects of different updated sets of parton distributions on the forward quantities,
namely CTEQ6L, CTEQ6L1 and MSTW.

The paper is organized as follows: in the next section we introduce a QCD-based eikonal model where the onset of the
dominance of semihard gluons in the interaction of high-energy hadrons is managed by the dynamical gluon mass. Within
this model we investigate a class of form factors which the spatial distribution of semihard gluons changes with energy
and introduce integral dispersion relations tailored to connect the real and imaginary parts of eikonals with this kind
of form factor.
In Section III we present the underlying physical picture of the infrared-finite QCD effective charge, and introduce
the elementary parton-parton cross sections connected to the gluon and quark dynamical masses.
In Section IV, motivated by the recent TOTEM measurements of cross sections at LHC, we perform a detailed analysis of
$pp$ and $\bar{p}p$ forward scattering data using our eikonal model, and obtain predictions for
$\sigma_{tot}^{pp,\bar{p}p}$ and $\rho^{pp,\bar{p}p}$ at Tevatron, CERN-LHC and cosmic-ray energies. The uncertainty on
these forward observables are inferred from the uncertainties associated with the dynamical mass scale and the parton
distribution functions at $\sqrt{s}=8$, 13, 14, 57 and 95 TeV. In Section V we draw our conclusions.

\section{The Dynamical Gluon Mass Model}

In the QCD-based (or ``mini-jet'') models the increase of the total cross sections is associated with semihard
scatterings of partons in the hadrons. These models incorporate soft and semihard processes in the treatment of high
energy hadron-hadron interactions using a formulation compatible with analyticity and unitarity constraints. In the
eikonal representation the total and inelastic cross sections, as well as the parameter $\rho$ (the ratio of the real
to imaginary part of the forward scattering amplitude), are given by
\begin{eqnarray}
\sigma_{tot}(s)   =  4\pi   \int_{_{0}}^{^{\infty}}   \!\!  b\,   db\,
[1-e^{-\chi_{_{R}}(s,b)}\cos \chi_{_{I}}(s,b)],
\label{eq01}
\end{eqnarray}
\begin{eqnarray}
\sigma_{inel}(s) = \sigma_{tot}(s) - \sigma_{el}(s) &=&  2\pi   \int_{_{0}}^{^{\infty}}   \!\!  b\,   db\,
G_{in}(s,b) \nonumber \\
 &=& 2\pi   \int_{_{0}}^{^{\infty}}   \!\!  b\,   db\,
[1-e^{-2\chi_{_{R}}(s,b)}],
\label{eq02}
\end{eqnarray}
\begin{eqnarray}
\rho(s) = \frac{-\int_{_{0}}^{^{\infty}}   \!\!  b\,  
db\, e^{-\chi_{_{R}}(s,b)}\sin \chi_{_{I}}(s,b)}{\int_{_{0}}^{^{\infty}}   \!\!  b\,  
db\,[1-e^{-\chi_{_{R}}(s,b)}\cos \chi_{_{I}}(s,b)]} ,
\label{eq03}
\end{eqnarray}
respectively, where $s$ is the square of the total center-of-mass energy, $b$ is the impact parameter, $G_{in}(s,b)$ is
the inelastic overlap function and
$\chi(s,b)=\textnormal{Re}\, \chi(s,b) + i\textnormal{Im}\, \chi(s,b) \equiv\chi_{_{R}}(s,b)+i\chi_{_{I}}(s,b)$
is the (complex) eikonal function. 

The unitarity of the $S$-matrix requires that the absorptive part of the elastic scattering amplitude receives
contributions from both the elastic and the inelastic channels. In impact parameter space this condition may be
written as
\begin{eqnarray}
2 \textnormal{Re} \Gamma (s,b) = |\Gamma (s,b)|^{2} + G_{in}(s,b) ,
\label{unitcond}
\end{eqnarray}
where $\Gamma (s,b)$ is the profile function, which describes the absorption resulting from the opening of inelastic
channels. It can be expressed by the inverse Fourier-Bessel transform of the elastic scattering
amplitude, $f(s,t)$,
\begin{eqnarray}
\Gamma(s,b) = -i \int_{0}^{\infty} d\sqrt{-t}\, \sqrt{-t}\, J_{0}(b\sqrt{-t})\, f(s,t),
\end{eqnarray}
where $t$ is the usual Mandelstam variable. The physical consequence of Eq. (\ref{unitcond}) is that no scattering
process can be uniquely inelastic, and thus the usual statement that the elastic amplitude results from shadow
scattering from the inelastic channels. In this picture the probability that neither hadron is broken up in a collision
at impact parameter $b$ is therefore given by $P(s,b)=e^{-2\chi_{_{R}}(s,b)}$.

We assume that the eikonal functions for $pp$ and $\bar{p}p$ scatterings are additive with respect to the soft and
semihard (SH) parton interactions in the hadron-hadron collision:
\begin{eqnarray}
\chi(s,b) = \chi_{_{soft}}(s,b) + \chi_{_{SH}}(s,b).
\end{eqnarray}

In the semihard limit of strong interactions hadron-hadron collisions can be treated as an incoherent sum of the
interactions among quarks and gluons. More specifically, the QCD cross section $\sigma_{_{QCD}}$ is obtained by
convoluting the cross sections $\hat{\sigma}$ for the QCD subprocesses with their associated parton distributions. It
follows from the QCD parton model that the eikonal term $\chi_{_{SH}}(s,b)$ can be factored as \cite{durand}
\begin{eqnarray}
\textnormal{Re}\,\chi_{_{SH}}(s,b) = \frac{1}{2}\, W_{\!\!_{SH}}(b)\,\sigma_{_{QCD}}(s),
\end{eqnarray}
where $W_{\!\!_{SH}}(b)$ is an overlap density for the partons at impact parameter space $b$,
\begin{eqnarray}
W_{\!\!_{SH}}(b) &=& \int d^{2}b'\, \rho_{A}(|{\bf b}-{\bf b}'|)\, \rho_{B}(b'),
\end{eqnarray}
and $\sigma_{_{QCD}}(s)$ is the usual QCD cross section 
\begin{eqnarray}
\sigma_{_{QCD}}(s) &=& \sum_{ij} \frac{1}{1+\delta_{ij}} \int_{0}^{1}\!\!dx_{1}
\int_{0}^{1}\!\!dx_{2} \int_{Q^{2}_{min}}^{\infty}\!\!d|\hat{t}|
\frac{d\hat{\sigma}_{ij}}{d|\hat{t}|}(\hat{s},\hat{t}) \nonumber \\
 &\times & f_{i/A}(x_{1},|\hat{t}|)f_{j/B}(x_{2},|\hat{t}|)\, \Theta \! \left( \frac{\hat{s}}{2} - |\hat{t}| \right),
\label{eq08}
\end{eqnarray}
with $|\hat{t}|\equiv Q^{2}$ and $i,j=q,\bar{q},g$. In the above expression the integration limits satisfy
$x_{1}x_{2}s > 2|\hat{t}| > 2Q^{2}_{min}$, where $Q^{2}_{min}$ is a minimal momentum transfer in the semihard scattering,
$\hat{s}$ and $\hat{t}$ are the Mandelstam variables of the parton-parton subsystem, and $x_{1}$ and $x_{2}$ are the
fractions of the momenta of the parent hadrons
$A$ and $B$ carried by the partons $i$ and $j$. The term $d\hat{\sigma}_{ij}/d|\hat{t}|$ is the differential cross
section for $ij$ scattering, and $f_{i/A}(x_{1},|\hat{t}|)$ ($f_{j/B}(x_{2},|\hat{t}|)$) is the usual parton $i$ ($j$)
distribution in the hadron $A$ ($B$).

The eikonal function is written in terms of even and odd eikonal parts connected by crossing symmetry. In the case of
the proton-proton ($pp$) and antiproton-proton ($\bar{p}p$) scatterings, this combination reads
$\chi_{pp}^{\bar{p}p}(s,b) = \chi^{+} (s,b) \pm \chi^{-} (s,b)$, with
$\chi^{+}(s,b) = \chi^{+}_{_{soft}}(s,b) + \chi^{+}_{_{SH}}(s,b)$ and
$\chi^{-}(s,b) = \chi^{-}_{_{soft}}(s,b) + \chi^{-}_{_{SH}}(s,b)$. However, in the QCD parton model
$\chi^{-}_{_{SH}}(s,b)$ decreases rapidly with increasing $s$, since the difference between $pp$ and $\bar{p}p$ cross
sections is due only to the different weighting of the quark-antiquark (valence) annihilation cross sections in the
two channels. Hence the crossing-odd eikonal $\chi^{-}(s,b)$ receives no contribution from semihard processes at high
energies. As a result, for our purposes it is sufficient to take $\chi_{_{SH}}(s,b)=\chi^{+}_{_{SH}}(s,b)$ and,
consequently, $\chi^{-}(s,b) = \chi^{-}_{_{soft}}(s,b)$. The connection between the real and imaginary parts of
$\chi^{+}(s,b)$ and $\chi^{-}(s,b)$, obtained by means of dispersion relations, will be discussed in the ensuing
sections.

\subsection{Energy-Dependent Form Factors}

For the overlap densities, the simplest hypothesis is to assume $W_{\!\!_{SH}}(b)$ be the same as
$W_{\!\!_{soft}}(b)$.
This prescription is not however true in the QCD parton model, since soft interactions are mainly related to
interactions among valence quarks, while semihard interactions are dominated by gluons. Moreover, it seems plausible
a scenario where quarks and gluons exhibit a somewhat different spatial distribution, since gluons are expected to be
distributed around the quarks. Furthermore, in contrast with gluons, quarks have electric charges, and the (matter)
distribution of the valence quarks can be associated in a reasonable way with the proton's charge distribution. As a
consequence, a commonly used choice for the soft overlap densities $W^{-}_{\!\!_{soft}}(b)$ and $W^{+}_{\!\!_{soft}}(b)$
comes from the charge dipole approximation to the form factors $G_{A}(k_{\perp})$ and $G_{B}(k_{\perp})$ of the
colliding hadrons $A$ and $B$, where
\begin{eqnarray}
W(b) &=& \int d^{2}b'\, \rho_{A}(|{\bf b}-{\bf b}'|)\, \rho_{B}(b')  \nonumber \\
 &=& \frac{1}{2\pi}\int_{0}^{\infty}dk_{\perp}\, k_{\perp}\, J_{0}(k_{\perp}b)\,G_{A}(k_{\perp})\,G_{B}(k_{\perp}),
\end{eqnarray}
and
 \begin{eqnarray}
G_{A}(k_{\perp})=G_{B}(k_{\perp})\equiv G_{dip}(k_{\perp};\mu)=\left( \frac{\mu^{2}}{k_{\perp}^{2}+\mu^{2}} \right)^{2}.
\end{eqnarray}
Here $\rho(b)$ is the parton density, which gives the probability density for finding a parton in the area $d^{2}b$
at impact parameter $b$. In terms of the form factor it is simply written as 
\begin{eqnarray}
\rho(b)=\frac{1}{(2\pi)^{2}}\int d^{2}k_{\perp}\, G(k_{\perp})e^{i{\bf k}_{\perp}\cdot {\bf b}}.
\end{eqnarray}
Thus, using the dipole form factor $G_{dip}(k_{\perp};\mu)$ one gets
\begin{eqnarray}
W^{+}_{\!\!_{soft}}(b;\mu^{+}_{_{soft}}) &=& \frac{1}{2\pi}\int_{0}^{\infty}dk_{\perp}\,
k_{\perp}\, J_{0}(k_{\perp}b)\,G_{dip}^{2}(k_{\perp};\mu^{+}_{_{soft}}) \nonumber \\
 &=& \frac{(\mu^{+}_{_{soft}})^{2}}{96\pi} (\mu^{+}_{_{soft}} b)^{3} K_{3}(\mu^{+}_{_{soft}} b),
\label{eq20}
\end{eqnarray}
where $K_{3}(x)$ is the modified Bessel function of second kind and $\mu^{+}_{_{soft}}$ is a free adjustable parameter
that accounts for the matter (valence quarks) distribution
inside the hadron. The $W(b;\mu)$ function is normalized so that $\int d^{2}b\, W(b;\mu) = 1$. In the same way, the odd
soft density is written as
\begin{eqnarray}
W^{-}_{\!\!_{soft}}(b;\mu^{-}_{_{soft}}) = \frac{(\mu^{-}_{_{soft}})^{2}}{96\pi}
(\mu^{-}_{_{soft}} b)^{3} K_{3}(\mu^{-}_{_{soft}} b),
\label{eq21}
\end{eqnarray} 
where $\mu^{-}_{_{soft}}\equiv 0.5$ GeV (its value is fixed since the odd eikonal
just accounts for the difference between $pp$ and $\bar{p}p$ channels at low energies).

In the case of semihard gluons, which dominate at high energy, we consider the possibility of a ``broadening'' of the
spatial distribution. Our assumption suggests an increase of the average gluon radius
when $\sqrt{s}$ increases. More important, it in fact does provide an excellent description of
$\sigma_{tot}^{pp,\bar{p}p}(s)$ and $\rho^{pp,\bar{p}p}(s)$ data, as shown in the next section, strongly suggesting an
energy dependence for the semihard overlap density. The way for introducing this effect can be paved by looking at
previous approaches, particularly in geometrical ones, in which the role of phenomenological energy-dependent form
factors is central \cite{formfactors1}. Our assumption, based on the QCD parton model, can be properly
implemented using two ansatz for the energy-dependent form factors, namely a monopole
\begin{eqnarray}
G^{(m)}_{\!\!_{SH}}(s,k_{\perp};\nu_{\!\!_{SH}})=\frac{\nu_{_{SH}}^{2}}{k_{\perp}^{2}+\nu_{_{SH}}^{2}},
\end{eqnarray}
and a dipole
\begin{eqnarray}
G^{(d)}_{\!\!_{SH}}(s,k_{\perp};\nu_{\!\!_{SH}})=\left( \frac{\nu_{_{SH}}^{2}}{k_{\perp}^{2}+\nu_{_{SH}}^{2}} \right)^{2},
\end{eqnarray}
where $\nu_{_{SH}}= \nu_{1}-\nu_{2}\ln ( \frac{s}{s_{0}} )$, with $\sqrt{s_{0}}\equiv 5$ GeV. Here $\nu_{1}$ and $\nu_{2}$
are constants to be fitted. In the case of the monopole the overlap density is
\begin{eqnarray}
W^{(m)}_{\!\!_{SH}}(s,b;\nu_{\!\!_{SH}}) &=& \frac{1}{2\pi}\int_{0}^{\infty}dk_{\perp}\, k_{\perp}\, J_{0}(k_{\perp}b)\,
[G^{(m)}_{\!\!_{SH}}(s,k_{\perp};\nu_{\!\!_{SH}})]^{2} \nonumber \\
 &=& \frac{\nu^{2}_{_{SH}}}{4\pi} (\nu_{_{SH}} b) K_{1}(\nu_{_{SH}} b),
\label{eq24}
\end{eqnarray}
where $K_{1}(x)$ is a modified Bessel function of second kind. In analogy with the Eq. (\ref{eq20}), in the case
of the dipole we are led to
\begin{eqnarray}
W^{(d)}_{\!\!_{SH}}(s,b;\nu_{\!\!_{SH}}) = \frac{\nu^{2}_{_{SH}}}{96\pi} (\nu_{_{SH}} b)^{3} K_{3}(\nu_{_{SH}} b).
\label{eq25}
\end{eqnarray}

Note that, as mentioned earlier, semihard interactions dominate at high energies. Thus we consider an
energy-dependence behavior for the spatial distribution exclusively in the case of $W_{\!\!_{SH}}(s,b)$. In this way,
the soft overlap densities $W^{-}_{\!\!_{soft}}(b)$ and $W^{+}_{\!\!_{soft}}(b)$ will emerge only from the ``static''
dipole form factor, i. e., from the Eqs. (\ref{eq20}) and (\ref{eq21}), whereas the semihard overlap density
$W_{\!\!_{SH}}(s,b)$ will be directly associated with Eqs. (\ref{eq24}) and (\ref{eq25}).
Moreover, in the semihard sector we have another form in which the eikonal can be factored into the QCD parton model,
since now $\chi_{_{SH}}(s,b) = \frac{1}{2}\, W_{\!\!_{SH}}(s,b)\,\sigma_{_{QCD}}(s)$.

\subsection{Integral Dispersion Relations and High-Energy Eikonal}

The analyticity of the scattering amplitude $f(s,t)$ leads to dispersion relations with crossing symmetry condition.
In the case of elastic processes in the forward direction ($t=0$), the crossing variable is the energy $E$ of the
incident particle in the laboratory frame \cite{blochcahn}. If ${\cal F}(E)$ is the analytic continuation of the
forward elastic scattering amplitude, $f(E,t=0)$, the $pp$ and $\bar{p}p$ forward amplitudes are the
limits of the analytic function ${\cal F}$ according to
\begin{eqnarray}
f^{\bar{p}p}_{pp}(E,t=0) = \lim_{\epsilon \to 0} {\cal F}(\mp E \mp i\epsilon, t=0) .
\end{eqnarray}

The Cauchy theorem implies that
\begin{eqnarray}
{\cal F}(E) = \frac{1}{2\pi i} \oint dE'\, \frac{{\cal F}(E')}{E'-E} 
\end{eqnarray}
and, after choosing an appropriate contour, the above expression can be written as
\begin{eqnarray}
{\cal F}(E) = \frac{1}{2\pi i} \left[ \int_{m}^{\infty} \!\! dE'\,
\frac{{\cal F}(E'+i\epsilon)-{\cal F}(E'-i\epsilon)}{E'-E} +\!\! 
\int_{-\infty}^{-m} \!\! dE'\, \frac{{\cal F}(E'+i\epsilon)-{\cal F}(E'-i\epsilon)}{E'-E}\right]\!\! ,
\end{eqnarray}
where $E=-m$ and $E=m$ are cuts on the real axis. For an even amplitude (${\cal F}={\cal F}^{+}$) we have
${\cal F}(E'+i\epsilon)={\cal F}(-E'-i\epsilon)$. Thus,
\begin{eqnarray}
{\cal F}^{+}(E) = \frac{1}{\pi } \int_{m}^{\infty} dE'\, \textnormal{Im}{\cal F}^{+}(E'+i\epsilon) \left[ 
\frac{1}{E'-E} + \frac{1}{E'+E} \right] ,
\end{eqnarray}
and the real and imaginary parts of ${f}^{+}(E)$ are connected by the dispersion relation
\begin{eqnarray}
\textnormal{Re}{f}^{+}(E) = \frac{2}{\pi }\, {\cal P}\!\! \int_{m}^{\infty} dE'
\left[ \frac{E'}{E'^{2}-E^{2}}  \right]
\textnormal{Im}{f}^{+}(E') ,
\label{jujuy001}
\end{eqnarray}
where ${\cal P}$ stands for the Cauchy principal-value integral. In our model the 
eikonals are written in terms of even and odd eikonal parts connected by crossing symmetry, namely
$\chi_{pp}^{\bar{p}p} = \chi^{+} \pm \chi^{-} $, where $\chi^{+} $ and $\chi^{-} $ are therefore
real analytic functions of $E$, i.e. they take real values on a real-axis segment, with
the same cut structure as $f^{+}$ and $f^{-}$, respectively. Hence, taking the
limit $E \gg m$ and changing the variable from $E$ to $s$, we find that the even eikonal also satisfies the reverse
dispersion relation
\begin{eqnarray}
\textnormal{Im}\,\chi^{+}(s,b) = -\frac{2s}{\pi}\, {\cal P}\!\! \int_{0}^{\infty}ds'\,
\frac{\textnormal{Re}\,\chi^{+}(s',b)}{s^{\prime 2}-s^{2}} .
\label{idr001}
\end{eqnarray}
Thus, integrating by parts:
\begin{eqnarray}
\textnormal{Im}\,\chi^{+}(s,b) &=& \lim_{\substack{\epsilon \to 0 \\ s''\to \infty}} -\frac{2s}{\pi} \left[
\int_{0}^{s-\epsilon}ds'\,\frac{\textnormal{Re}\,\chi^{+}(s',b)}{s^{\prime 2}-s^{2}}    +
\int_{s+\epsilon}^{s''}ds'\,\frac{\textnormal{Re}\,\chi^{+}(s',b)}{s^{\prime 2}-s^{2}}
\right] \nonumber \\
 &=& \lim_{s''\to \infty} \frac{1}{\pi} \left[ \textnormal{Re}\,\chi^{+}(s'',b)\ln \left( \frac{s''+s}{s''-s} \right) -
\int_{0}^{\infty}ds'\,\ln \left( \frac{s'+s}{|s'-s|} \right) \frac{d\,\textnormal{Re}\,\chi^{+}(s',b)}{ds'}
\right] \nonumber \\
 &=& -\frac{1}{\pi} \int_{0}^{\infty}ds'\,\ln \left( \frac{s'+s}{|s'-s|} \right) \frac{d\,
\textnormal{Re}\,\chi^{+}(s',b)}{ds'} ,
\label{disrel001}
\end{eqnarray}
where in the last step we have observed that the first term vanishes in the limit $s''\to \infty$. Applying this
dispersion relation to
$\textnormal{Re}\,\chi_{_{SH}}(s,b)=\textnormal{Re}\,\chi^{+}_{_{SH}}(s,b)=\frac{1}{2}W_{\!\!_{SH}}(s,b)\sigma_{_{QCD}}(s)$,
we get
\begin{eqnarray}
\textnormal{Im}\,\chi_{_{SH}}(s,b) &=& -\frac{1}{2\pi}\, \int_{0}^{\infty} ds'\, 
\ln \left( \frac{s'+s}{|s'-s|} \right) \left[ \sigma_{_{QCD}}(s')\,
\frac{dW_{\!\!_{SH}}(s',b)}{ds'} \right] \nonumber \\
& & -\frac{1}{2\pi}\, \int_{0}^{\infty} ds'\, 
\ln \left( \frac{s'+s}{|s'-s|} \right) \left[ W_{\!\!_{SH}}(s',b)\, \frac{d\sigma_{_{QCD}}(s')}{ds'} \right] .
\label{jujuy2}
\end{eqnarray}
The second integral on the right side involves the derivative of the QCD cross section $\sigma_{_{QCD}}(s')$. We should
at this point note that the $s'$ dependence in $\frac{d\hat{\sigma}_{ij}}{d|\hat{t}|}$ terms can be ignored, since
their derivatives are of order $1/s^{\prime 2}$. In this way, the only energy dependence appears in the Heaviside function
$\Theta$ (see (\ref{eq08})), in which
\begin{eqnarray}
\frac{d}{ds'}\, \Theta \! \left( \frac{\hat{s}'}{2}-|\hat{t}| \right) = 
\frac{d}{ds'}\, \Theta \! \left( s' -\frac{2|\hat{t}|}{x_{1}x_{2}} \right) =
\delta \! \left( s' -\frac{2|\hat{t}|}{x_{1}x_{2}} \right) .
\label{delta001}
\end{eqnarray}
The $\delta$-function removes the integration over $ds'$, thus, the second integral can be expressed as
\begin{eqnarray}
I_{2}(s,b) &=& -\frac{1}{2\pi}\, \int_{0}^{\infty} ds'\, \ln\left( \frac{s'+s}{|s'-s|} 
\right) W_{\!\!_{SH}}(s',b)\, \frac{d\sigma_{_{QCD}}(s')}{ds'} \nonumber \\
 &=& -\frac{1}{2\pi}\sum_{ij} \frac{1}{1+\delta_{ij}} \, W_{\!\!_{SH}}\!\left(\frac{2|\hat{t}|}{x_{1}x_{2}},b\right) 
\int_{0}^{1}\!\!dx_{1}
\int_{0}^{1}\!\!dx_{2} \int_{Q^{2}_{min}}^{\infty}\!\!\!\!d|\hat{t}|\,
\frac{d\hat{\sigma}_{ij}}{d|\hat{t}|}(\hat{s},\hat{t}) \nonumber \\ 
 &\times& f_{i/A}(x_{1},|\hat{t}|)f_{j/B}(x_{2},|\hat{t}|)
\ln \left( \frac{\hat{s}/2+|\hat{t}|}{\hat{s}/2-|\hat{t}|}\right) \Theta \! \left( \frac{\hat{s}}{2} - |\hat{t}| 
\right) .
\label{eq11}
\end{eqnarray}
The energy-dependent form factor $W_{\!\!_{SH}}(s,b)$ can have a monopole or a dipole form, namely
$W^{(m)}_{\!\!_{SH}}(s,b;\nu_{\!\!_{SH}})$ or $W^{(d)}_{\!\!_{SH}}(s,b;\nu_{\!\!_{SH}})$ (see eqs. (\ref{eq24}) and (\ref{eq25})).
In the case of a monopole form, the first integral on the right side of (\ref{jujuy2}) can be rewritten as
\begin{eqnarray}
I^{(m)}_{1}(s,b) &=& -\frac{1}{2\pi}\, \int_{0}^{\infty} ds'\, \ln\left( \frac{s'+s}{|s'-s|} \right) 
\sigma_{_{QCD}}(s')\,
\frac{dW^{(m)}_{\!\!_{SH}}(s',b;\nu_{\!\!_{SH}})}{ds'} \nonumber \\
 &=& -\frac{b}{8\pi^{2}}\sum_{ij} \frac{1}{1+\delta_{ij}} \int_{0}^{\infty} \frac{ds'}{s'}\, 
\ln\left( \frac{s'+s}{|s'-s|} \right) 
\int_{0}^{1}\!\!dx_{1} \int_{0}^{1}\!\!dx_{2} \int_{Q^{2}_{min}}^{\infty}\!\!\!\!d|\hat{t}|\,
\frac{d\hat{\sigma}_{ij}}{d|\hat{t}|}(\hat{s}',\hat{t}) \nonumber \\ 
 &\times& f_{i/A}(x_{1},|\hat{t}|)f_{j/B}(x_{2},|\hat{t}|)\left[ b\nu_{2}\nu_{_{SH}}^{3}K_{0}(\nu_{_{SH}} b)-
2\nu_{2}\nu_{_{SH}}^{2}K_{1}(\nu_{_{SH}} b)  \right] 
\Theta \! \left( \frac{\hat{s}'}{2} - |\hat{t}| \right) ; \nonumber
\label{eq12a}
\end{eqnarray}
in the case of a dipole we get
\begin{eqnarray}
I^{(d)}_{1}(s,b) &=& -\frac{1}{2\pi}\, \int_{0}^{\infty} ds'\, \ln\left( \frac{s'+s}{|s'-s|} \right) 
\sigma_{_{QCD}}(s')\,
\frac{dW^{(d)}_{\!\!_{SH}}(s',b;\nu_{\!\!_{SH}})}{ds'} \nonumber \\
&=& -\frac{b^{3}}{192\pi^{2}}\sum_{ij} \frac{1}{1+\delta_{ij}} \int_{0}^{\infty} \frac{ds'}{s'}\, \ln\left( 
\frac{s'+s}{|s'-s|} \right) 
\int_{0}^{1}\!\!dx_{1} \int_{0}^{1}\!\!dx_{2} \int_{Q^{2}_{min}}^{\infty}\!\!\!\!d|\hat{t}|\,
\frac{d\hat{\sigma}_{ij}}{d|\hat{t}|}(\hat{s}',\hat{t}) \nonumber \\ 
 &\times& f_{i/A}(x_{1},|\hat{t}|)f_{j/B}(x_{2},|\hat{t}|)\left[ b\nu_{2}\nu_{_{SH}}^{5}K_{2}(\nu_{_{SH}} b)-
2\nu_{2}\nu_{_{SH}}^{4}K_{3}(\nu_{_{SH}} b)  \right] 
\Theta \! \left( \frac{\hat{s}'}{2} - |\hat{t}| \right) . \nonumber
\label{eq12b}
\end{eqnarray}

The soft eikonal is needed only to describe the lower-energy forward data, since the main contribution to the
asymptotic behavior of hadron-hadron total cross section comes from parton-parton semihard collisions. Therefore it
is enough to build an instrumental parametrization for the soft eikonal with terms dictated by the Regge
phenomenology \cite{regge001}. For the even part of the soft eikonal we therefore take
\begin{eqnarray}
\chi^{+}_{_{soft}}(s,b) = \frac{1}{2}\, W^{+}_{\!\!_{soft}}(b;\mu^{+}_{_{soft}})\, \left[ A' +\frac{B'}{(s/s_{0})^{\gamma}}\, e^{i\pi\gamma/2}
-i C'\left[ \ln\left(\frac{s}{s_{0}}\right) -i\frac{\pi}{2} \right] \right] ,
\label{soft01}
\end{eqnarray}
where $\sqrt{s_{0}}\equiv 5$ GeV and $A'$, $B'$, $C'$, $\gamma$ and $\mu^{+}_{_{soft}}$ are fitting parameters. The phase
factor $e^{i\pi\gamma/2}$, which ensures the correct analyticity properties of the amplitude, is a result of the integral
dispersion relation (\ref{idr001}).

The odd eikonal $\chi^{-}(s,b)$, that  accounts for the difference between $pp$ and $\bar{p}p$ channels and vanishes
at high energy, is given by
\begin{eqnarray}
\chi^{-}(s,b) &=& \frac{1}{2}\, W^{-}_{\!\!_{soft}}(b;\mu^{-}_{_{soft}})\,D'\, \frac{e^{-i\pi/4}}{\sqrt{s/s_{0}}},
\label{softminus}
\end{eqnarray}
where $D'$, the strength of the odd term, is also a fitting parameter. The expression (\ref{softminus}) was written
with its correct analyticity property, since the phase factor $e^{-i\pi/4}$ is a result of the dispersion relation
(valid at $s\gg m$)
\begin{eqnarray}
\textnormal{Im}\,\chi^{-}(s,b) = -\frac{2s^{2}}{\pi}\, {\cal P}\!\! \int_{0}^{\infty}ds'\,
\frac{\textnormal{Re}\,\chi^{-}(s',b)}{s'(s^{\prime 2}-s^{2})} .
\label{idr002}
\end{eqnarray}

\section{Infrared mass scale and the role of gluons}

The calculation of the QCD cross section $\sigma_{_{QCD}}(s)$ implies the sum over all possible parton types,
but it is sufficiently accurate for our purposes to fix the number
of flavors $n_{f}=4$ and keep only the gluon $g$ and the quarks $u$, $d$, $s$ and $c$. As a matter of fact
$\textnormal{Re}\,\chi_{_{SH}}(s,b)$ and $\textnormal{Im}\,\chi_{_{SH}}(s,b)$ have to be determined taking into account
all the heavy quarks, where each heavy quark $h=c,b,t$ with mass $M_{h}$ is effectively decoupled from physical
cross sections at momenta scales below the thresholds $Q_{h}=M_{h}$, being $n_{f}$ an increasing function of $Q_{h}$.
However, our numerical results show that the contributions of the quarks $b$ and $t$ to $\chi_{_{SH}}(s,b)$
are very small indeed. In fact even the charm contribution is tiny, but was included only for
high-precision purposes. Hence there is no fundamental role for heavy quarks ($m_{q} \approx M_{h}$, $h=c,b,t$) in our
analysis, and this result can be understood as follows: heavy quarks are produced (perturbatively) from the splitting
of gluons into $\bar{h}h$ pairs at energies above the thresholds $Q_{h}=M_{h}$. At sufficiently small $x$, the ratio
of the heavy-quark parton distribution function,
$h(x,Q^{2})$, to the gluon one, $g(x,Q^{2})$, is \cite{tung001}
\begin{eqnarray}
\frac{h(x,Q^{2})}{g(x,Q^{2})} \sim \frac{\alpha_{s}(Q^{2})}{2\pi}\ln \left( \frac{Q^{2}}{M^{2}_{h}} \right) ,
\label{heavy009} 
\end{eqnarray}
where $h(x,Q^{2})=0$ at $Q=M_{h}$. However, the angular dependence of the dominant subprocesses in (\ref{eq08})
are very similar and all dominated by the $t$-channel angular distribution and, as consequence, the
parton-parton differential cross sections vary essentially as $d\hat{\sigma}_{ij}/d|\hat{t}| \sim 1/Q^{4}$.
Hence the effects of distribution functions as well as current masses of heavy quarks on $\sigma_{_{QCD}}(s)$
are absolutely negligible.

In order to obtain $\chi_{_{SH}}(s,b)$ we select parton-parton scattering processes containing at least one gluon in
the initial state. The reason for this choice comes from the behavior of the partonic splitting dictated by the DGLAP
evolution equations at leading-order \cite{dglap}, in which the gluon splitting functions $P_{gq}\to \frac{4}{3z}$
and $P_{gg}\to \frac{6}{z}$ are singular as $z\to 0$. As a result, the gluon distribution becomes very large as
$x\to 0$ (in the convolution integrals $z<x$), and its role in the evolution of parton distributions becomes central.
Thus, we select the following
processes in the calculation of $\chi_{_{SH}}(s,b)$: $gg\to gg$ (gluon-gluon scattering), $qg\to qg$ and
$\bar{q}g\to \bar{q}g$ (quark-gluon scattering), and $gg\to \bar{q}q$ (gluon fusion into a quark pair). The
gluon-gluon and quark-gluon scattering processes in fact dominate at high energies. For example, at $\sqrt{s}=7$ TeV
and with $Q_{min}=1.3$ GeV, their relative contribution to the cross section $\sigma_{_{QCD}}(s)$ is around 98,84\%
(98,66\%) for the CTEQ6L (MSTW) set of parton distributions. The relative contribution of the process
$gg\to \bar{q}q$ is tiny; nevertheless it was included for completeness.

These elementary processes are plagued by infrared divergences, which have to be regularized by means of some cutoff
procedure. One natural regulator for these infrared divergences was introduced some time ago \cite{cornwall}, and has
become an important ingredient of our eikonal models \cite{luna001,luna009}. It is based on the increasing evidence
that QCD develops an effective, momentum-dependent mass for the gluons, while preserving the local $SU(3)_{c}$
invariance of the theory. This {\it dynamical} mass $M_{g}(Q^{2})$ introduces a natural scale that, in principle,
sets up a threshold for gluons to pop up from the vacuum \cite{soni001}. Moreover, it is intrinsically linked to an
infrared-finite QCD effective charge $\bar{\alpha}_{s}(Q^{2})$, therefore being the natural infrared regulator in our
eikonal model.

Since the gluon mass generation is a purely dynamical effect, the formal tool for tackling this non-perturbative
phenomenon, in the continuum, is provided by the Schwinger-Dyson equations \cite{sde001}. These equations constitute
an infinite set of coupled non-linear integral equations governing the dynamics of all QCD Green's functions. The
functional forms of $M_{g}$ and $\bar{\alpha}_{s}$, obtained by Cornwall through the use of the pinch technique in
order to derive a gauge invariant Schwinger-Dyson equation for the gluon propagator and the triple gluon vertex, are
given by \cite{cornwall}
\begin{eqnarray} 
\bar{\alpha}_{s} (Q^{2})= \frac{4\pi}{\beta_0 \ln\left[
(Q^{2} + 4M_g^2(Q^{2}) )/\Lambda^2 \right]}, 
\label{eq26}
\end{eqnarray}
\begin{eqnarray}
M^2_g(Q^{2}) =m_g^2 \left[\frac{ \ln
\left(\frac{Q^{2}+4{m_g}^2}{\Lambda ^2}\right) } {
\ln\left(\frac{4{m_g}^2}{\Lambda ^2}\right) }\right]^{- 12/11} ,
\label{mdyna} 
\end{eqnarray}
where $\Lambda$($\equiv\Lambda_{QCD}$)  is the QCD scale parameter, $\beta_0 =  11- \frac{2}{3}n_f$,
and $m_{g}$ is an infrared mass scale to be adjusted in order to provide reliable results concerning calculations of
strongly interacting processes. As mentioned in the earlier section, the existence of the gluon mass scale $m_{g}$ is
strongly supported by QCD lattice simulations and phenomenological results, and its value is typically found to be of
the order $m_{g}=500\pm200$ MeV. Note that in the limit $Q^{2} \gg \Lambda^{2}$ the dynamical mass
$M_{g}(Q^{2})$ vanishes, and the effective charge matches with the one-loop perturbative QCD coupling
$\alpha^{pQCD}_{s}(Q^{2})$. It means that the asymptotic ultraviolet behavior of the LO running coupling, obtained
from the renormalization group equation perturbation theory, is reproduced in solutions of Schwinger-Dyson equations,
\begin{eqnarray}
\bar{\alpha}_{s} (Q^2 \gg \Lambda^2) \sim
\frac{4\pi}{\beta_{0} \ln \left( \frac{Q^2 }{\Lambda^2} \right) } = \alpha^{pQCD}_{s} (Q^2) ,
\label{eq41}
\end{eqnarray}
provided only that the truncation method employed in the analysis preserves the multiplicative renormalizability
\cite{luna009}. However, in the infrared region, the coupling $\alpha^{pQCD}_{s}(Q^{2})$ has Landau
singularities on the space-like semiaxis $0 \leq Q^{2} \leq \Lambda ^2$, i.e. has a nonholomorphic (singular) behavior
at low $Q^{2}$ (for a recent review, see \cite{stefanis001}). This problem has been faced in the past years with
analytic versions of QCD whose coupling $\alpha_{s}(Q^{2})$ is holomorphic (analytic) in the entire complex plane
except the time-like axis ($Q^{2}<0$) \cite{cvetic007}.
Our effective charge $\bar{\alpha}_{s} (Q^{2})$, on the other hand, shows the existence of an infrared
fixed point as $Q^2\rightarrow  0$, i.e. the dynamical mass term tames the Landau pole and $\bar{\alpha}_{s}$
freezes at a finite value in the infrared limit. Thus, providing that the gluon mass scale is set larger than half of
the QCD scale parameter, namely $m_{g}/\Lambda >1/2$, the analyticity of $\bar{\alpha}_{s} (Q^{2})$ is preserved.
This ratio is also phenomenologically determined \cite{luna001,luna009,durao} and
typically lies in the interval $m_{g}/\Lambda \in [1.1, 2]$.
Moreover, as recently pointed out by Cveti\v{c} \cite{cvetic004}, evaluation of renormalization scale-invariant
spacelike quantities at low $Q^{2}$, in terms of infrared freezing couplings, can be done as a truncated series in
derivative of the coupling with respect to the logarithm of $Q^{2}$, which in turn exhibit significant better
convergence
properties.

Hence, taking into account the mechanism of dynamical mass generation in QCD, the required
parton-parton cross sections for calculating $\sigma_{_{QCD}}(s)$ are given by
\begin{eqnarray}
\frac{d\hat{\sigma}}{d\hat{t}}(gg\to gg)=\frac{9\pi\bar{\alpha}^{2}_{s}}{2\hat{s}^{2}}\left(3 -
\frac{\hat{t}\hat{u}}{\hat{s}^{2}}-\frac{\hat{s}\hat{u}}{\hat{t}^{2}}-\frac{\hat{t}\hat{s}}{\hat{u}^{2}} \right) ,
\label{pp001}
\end{eqnarray}
\begin{eqnarray}
\frac{d\hat{\sigma}}{d\hat{t}}(qg\to qg)=\frac{\pi\bar{\alpha}^{2}_{s}}{\hat{s}^{2}}\, (\hat{s}^{2}+\hat{u}^{2}) \left(
\frac{1}{\hat{t}^{2}}-\frac{4}{9\hat{s}\hat{u}} \right) ,
\label{pp002}
\end{eqnarray}
\begin{eqnarray}
\frac{d\hat{\sigma}}{d\hat{t}}(gg\to \bar{q}q)=\frac{3\pi\bar{\alpha}^{2}_{s}}{8\hat{s}^{2}}\,
(\hat{t}^{2}+\hat{u}^{2}) \left( \frac{4}{9\hat{t}\hat{u}}-\frac{1}{\hat{s}^{2}} \right) .
\label{pp003}
\end{eqnarray}
We call attention to the fact that, in the limit of large enough $Q^{2}$, the expressions (\ref{pp001}),
(\ref{pp002}) and (\ref{pp003}) reproduce their pQCD counterparts. In these expressions the kinematic constraints
under consideration are given by $\hat{s}+\hat{t}+\hat{u}=4M_{g}^{2}(Q^{2})$ in the case of gluon-gluon scattering,
and $\hat{s}+\hat{t}+\hat{u}=2M_{g}^{2}(Q^{2})+2M_{q}^{2}(Q^{2})$ in the cases of quark-gluon and gluon fusion into a
quark pair. Here $M_{q}(Q^{2})$ is the dynamical quark mass,
\begin{eqnarray}
M_{q}(Q^{2}) = \frac{m_{q}^{3}}{Q^{2}+m_{q}^{2}} ,
\label{quark001}
\end{eqnarray}
which assumes a non-zero infrared mass scale $m_{q}$, to be phenomenologically adjusted. Notice that the effective mass
for quarks is a sum of the dynamical mass and the running one. However, as discussed above, only the
contributions of lighter quarks are relevant in calculating the QCD cross section $\sigma_{_{QCD}}(s)$ and as a result
the effective mass behavior is dominated by the dynamical part. 
The expression (\ref{quark001}),
which decreases rapidly with increasing $Q$, is the simplest ansatz for a dynamical quark mass in agreement with the
operator product expansion (OPE) \cite{ope1,ope2,ope3,ope}. According to the OPE the dynamical mass is a function of
the quark condensate $\langle {\bar \psi}\psi \rangle$. More specifically,
$M_{q}(P^{2}) \propto \langle {\bar \psi}\psi \rangle/P^{2}$, where $P^{2}=-p^{2}$ is the momentum in Euclidean space.
The quark mass scale $m_{q}$ can be related to the quark condensate
($\langle {\bar \psi}\psi \rangle \propto m_{q}^{3}$ by dimensional considerations) and general constraints are
satisfied for $m_{q} \in [100,250]$ MeV. The simple power-law behavior of $M_{q}(Q^{2})$ is finally obtained by
introducing the factor $m_{q}^{2}$ in the denominator in order to get the right infrared limit
$M^{2}_{q}(Q^{2}\to 0) = m_{q}^{2}$.

\section{Results}

First, in order to determine the model parameters, we fix $n_f = 4$ and set the values of the gluon and quark mass
scales to $m_{g} = 400$ MeV and $m_{q} = 250$ MeV, respectively. These choices for the mass scales are not only
consistent to our LO procedures, but are also the ones usually obtained in other calculations of strongly interacting
processes \cite{luna001,luna009,luna010,luna2015}. Next, we carry out a global fit to high-energy forward $pp$ and
$\bar{p}p$ scattering data above $\sqrt{s} = 10$ GeV, namely the total cross section $\sigma_{tot}^{pp,\bar{p}p}$ and
the ratio of the real to imaginary part of the forward scattering amplitude $\rho^{pp,\bar{p}p}$. We use data sets
compiled and analyzed by the Particle Data Group \cite{PDG} as well as the recent data at LHC from the TOTEM
Collaboration, with the statistic and systematic errors added in quadrature. The TOTEM dataset includes the first and
second measurements of the total $pp$ cross section at $\sqrt{s}=7$ TeV, $\sigma_{tot}^{pp}=98.3\pm 2.8$
\cite{TOTEM001} and $\sigma_{tot}^{pp}=98.58\pm2,23$ \cite{TOTEM002} (both using the optical theorem together with
the luminosity provided by the CMS \cite{CMSlum}), the luminosity-independent measurement at $\sqrt{s}=7$ TeV,
$\sigma_{tot}^{pp}=98.0\pm2.5$ \cite{TOTEM003}, the $\rho$-independent measurement at $\sqrt{s}=7$ TeV,
$\sigma_{tot}^{pp}=99.1\pm4.3$ \cite{TOTEM003}, and the luminosity-independent measurement at $\sqrt{s}=8$ TeV,
$\sigma_{tot}^{pp}=101.7\pm2.9$ \cite{TOTEM004}.
We include in the dataset the first estimate for the $\rho$ parameter made by the TOTEM Collaboration in their 
$\rho$-independent measurement at $\sqrt{s}=7$ TeV, namely $\rho^{pp}=0.145\pm0.091$ \cite{TOTEM003}. In all the fits
performed in this paper we use a $\chi^{2}$ fitting procedure, assuming an interval $\chi^{2}-\chi_{min}^{2}$
corresponding, in the case of normal errors, to the projection of the $\chi^{2}$ hypersurface containing 90\% of
probability. In our model (8 fitting parameters) this corresponds to the interval $\chi^{2}-\chi_{min}^{2}=13.36$.

In our analysis we have investigated the effects of some updated sets of PDFs on the high-energy cross sections. In
performing the fits one uses tree-level formulas for the parton-parton cross sections. In this way we have to choose
parton distributions functions evolved with LO splitting functions, as in case of LO sets CTEQ6L, CTEQ6L1 and MSTW.
For the coupling $\alpha_{s}(Q^{2})$ it is usual to use either the LO formula for formal consistency or even the NLO
one. In the specific case of CTEQ distributions \cite{cteq6}, the CTEQ6L1 uses LO formula for $\alpha_{s}(Q^{2})$ with
$\Lambda^{(4flavor)}_{CTEQ6L1}=215$ MeV, whereas CTEQ6L uses NLO formula for $\alpha_{s}(Q^{2})$ with
$\alpha_{s}(M_{Z})=0.118$, consistent with the value $\Lambda^{(4flavor)}_{CTEQ6L}=326$ MeV. Since the dynamical mass
$M_{g}(Q^{2})$ practically vanishes at scales where four flavors are active, we choose these same values of
$\Lambda^{(4flavor)}$ in our effective charges $\bar{\alpha}_{s}^{LO}(Q^{2})$ and $\bar{\alpha}_{s}^{N\!LO}(Q^{2})$, where
$\bar{\alpha}_{s}^{LO}$ is given by the expression (\ref{eq26}) whereas $\bar{\alpha}_{s}^{N\!LO}(Q^{2})$
is given by \cite{luna009}
\begin{eqnarray}
\bar{\alpha}_{s}^{N\!LO}(Q^{2}) = \frac{4\pi}{\beta_{0}\ln\left(\frac{Q^{2} +
4M^{2}_{g}(Q^{2})}{\Lambda^{2}}\right)}\left[1-\frac{\beta_{1}}{\beta_{0}^{2}}\frac{\ln\left(\ln\left(\frac{Q^{2} +
4M^{2}_{g}(Q^{2})}{\Lambda^{2}}\right)\right)}{\ln\left(\frac{Q^{2} + 4M^{2}_{g}(Q^{2})}{\Lambda^{2}}\right)} \right],
\label{ansatz2}
\end{eqnarray}
where $\beta_{1} =102 -\frac{38}{3}n_{f}$ and $\Lambda = \Lambda^{(4flavor)}_{CTEQ6L}$. This NLO non-perturbative
coupling is built by saturating the two-loop perturbative strong coupling $\alpha_{s}^{N\!LO}$, that is, by
introducing the replacement
$\alpha_{s}^{N\!LO}(Q^{2}) \to \bar{\alpha}_{s}^{N\!LO}(Q^{2}) = \alpha_{s}^{N\!LO}(Q^{2} + 4M^{2}_{g}(Q^{2}))$ into the
perturbative result. Note that we are using the same dynamical mass $M^{2}_{g}(Q^{2})$ expression for both LO and NLO
couplings, since the results from the reference \cite{luna009} give support to the statement that the dynamical mass
scale $m_{g}$ is not strongly dependent on the perturbation order.
The MSTW set uses an alternative definition of $\alpha_{s}$, where the renormalization group equation for $\alpha_{s}$
is truncated at the appropriate order and solved starting from an initial value $\alpha_{s}(Q_{0}^{2})$. This input
value is one of their fit parameters and replaces the $\Lambda$ parameter \cite{mstw}. In the usual
matching-prescription scheme the behavior of $\alpha_{s_{MSTW}}(Q^{2})$ can be properly reproduced from the
choice $\Lambda^{(4flavor)}_{MSTW}\sim 319$ MeV. 

The values of the fitted parameters are given in Tables 1 and 2. In Table 1 (2) we show the values of the parameters
in the case of a monopole (dipole) form factor in the semihard sector. The $\chi^{2}/DOF$ for all fits was obtained
for 154 degrees of freedom. The sensitivity of the $\chi^{2}/DOF$ to the cutoff $Q_{min}$ is shown in Fig. 1. We
observe that the $\chi^{2}/DOF$ is not very sensitive to $Q_{min}$ in the interval $[1.0, 1.5]$ GeV for all PDFs we
have considered. The results of the fits to $\sigma_{tot}$ and $\rho$ for both $pp$ and $\bar{p}p$ channels are
displayed in Figs. 2, 3, 4 and 5, together with the experimental data. In Figure 6 we show the theoretical predictions
for the $pp$ cross sections at cosmic-ray energies; the comparison of the curves with the AUGER experimental datum
at $\sqrt{s} = 57$ TeV \cite{auger} and the Telescope Array datum at $\sqrt{s} = 95$ TeV \cite{ta001} shows good
agreement.
The curves depicted in Figs. 2-6 were all calculated using the cutoff $Q_{min}=1.3$ GeV, the value of the CTEQ6 fixed
initial scale $Q_{0}$. In the case of MSTW set the slightly lower value $Q_{0}\equiv 1$ GeV is adopted, and the
condition $Q_{min}\ge Q_{0}$ is always satisfied in our analysis.
In the case of fits using the CTEQ6 set, calculations in the region $Q_{min}< Q \leq Q_{0}$ were carried out
with PDFs fixed at the scale $Q=Q_{0}=1.3$ GeV.
In Table III we show the theoretical predictions for the forward scattering quantities $\sigma_{tot}^{pp,\bar{p}p}$ and
$\rho^{pp,\bar{p}p}$ using different sets of parton distributions.

\section{Conclusions}

In this paper we have studied infrared contributions to semihard parton-parton interactions by considering LO and NLO
effective QCD charges with finite infrared behavior. We have investigated $pp$ and $\bar{p}p$ scattering in the LHC
energy region with the assumption that the observed increase of hadron-hadron total cross sections arises exclusively
from these semihard interactions.
In the calculation of $\sigma_{tot}^{pp,\bar{p}p}$ and $\rho^{pp,\bar{p}p}$ we have investigated the behavior of the
forward amplitude for a range of different cutoffs and parton distribution functions, namely CTEQ6L, CTEQ6L1 and
MSTW, and considered the phenomenological implications of a class of energy-dependent form factors for semihard
partons. We introduce integral dispersion relations specially tailored to connect the real and imaginary parts of
eikonals with energy-dependent form factors.
In our analysis we have included the recent data at LHC from the TOTEM Collaboration. We have paid attention
to the sensitivity of the $\chi^{2}/DOF$ to the cutoff $Q_{min}$, which restrict the parton-parton processes to
semihard interactions. Our results show that very good descriptions of $\sigma_{tot}^{pp,\bar{p}p}$ and
$\rho^{pp,\bar{p}p}$ data are obtained by constraining the cutoff to the interval $1.0 \le Q_{min}\lesssim 1.5$ GeV.
The $\chi^{2}/DOF$ for the best global fits lies in the range [1.05, 1.06] for 154 degrees of freedom. This good
statistical result shows that our eikonal model, where non-perturbative effects are naturally included via a QCD
effective charge, is well suited for detailed predictions of the forward quantities to be measured at higher energies.
In fact our predictions for $pp$ total cross section are statistically compatible with the AUGER result at
$\sqrt{s}=57$ TeV, namely
$\sigma^{pp}_{tot}=[133\pm 13(\textnormal{stat})^{+17}_{-20}(\textnormal{syst})\pm 16(\textnormal{Glauber})]$ mb
\cite{auger}, as well as with the Telescope Array result at $\sqrt{s}=95$ TeV, namely
$\sigma^{pp}_{tot}=[170^{+48}_{-44}(\textnormal{stat})^{+17}_{-19}(\textnormal{syst})]$ mb \cite{ta001}. However it is
worth noting that both results are model dependent, since they come from the conversion of the proton-air production
cross section via a Glauber calculation. Moreover, as stressed by AUGER group, the total uncertainty of converting
the proton-air to $pp$ cross section may be larger than the published. Clearly new results from AUGER and Telescope
Array at higher energies would be extremely informative.

The uncertainty in our theoretical predictions for the forward observables at $\sqrt{s}=8$, 13, 14, 57 and 95 TeV
(Table III) have been estimated by varying the gluon mass scale within a typical uncertainty $\delta m_{g}$ while
keeping all other model parameters constant, and by exploring the uncertainties of parton distributions on production
cross sections.
This procedure does not determines the {\it formal} uncertainty in $\sigma_{tot}$
and $\rho$, since the variance-covariance matrix method, necessary for proving this quantity, was not employed.
However, at high energies the forward observables are dominated by semihard interactions represented by the eikonal
term $\chi_{_{SH}}(s,b)$, which depends only on 3 parameters, namely $\nu_{1}$, $\nu_{2}$ and $m_{g}$. In all $\chi^{2}$
analyses we have observed that the correlation coefficients of these parameters are very small. Moreover, the values
of $\sigma_{tot}$ and $\rho$ are actually more sensitive to the gluon mass scale than to variations of others
parameters of the model.
A reliable estimate of $\delta m_{g}$, namely around 7.1\% of the value of $m_{g}$, was obtained from the
analysis of the proton structure function $F_{2}(x,Q^{2})$ at small-$x$ \cite{luna009}.
Hence in our case, where $m_{g}$ was set at 400 MeV, the gluon mass uncertainty is $\delta m_{g}\sim 28$ GeV. 
In order to estimate the uncertainty of parton distributions on the forward predictions we simply adopt the
conservative stance that the PDFs uncertainties on the total cross sections are of the same order of magnitude as the
uncertainties on the production cross sections of the $W$ and $Z$ bosons at the LHC. The uncertainties on the
production cross sections are estimated to be $\pm5\%$ by the CTEQ group \cite{cteq6,cteq6a}. So, in sum, the total
uncertainty of our theoretical predictions is obtained from the quadrature sum of the uncertainties coming
from the gluon mass uncertainty $\delta m_{g}$ and the parton distributions.

In the semihard sector we have considered a new class of form factors in which the average gluon radius increases
with $\sqrt{s}$. With this assumption we have obtained another form in which the eikonal can be factored into the QCD
parton model, namely
$\textnormal{Re}\,\chi_{_{SH}}(s,b) = \frac{1}{2}\, W_{\!\!_{SH}}(s,b)\,\textnormal{Re}\,\sigma_{_{QCD}}(s)$. The imaginary
part of this {\it semi-factorizable} eikonal was obtained by means of appropriate integral dispersion relations
which take into account eikonals with energy-dependent form factors. Although these dispersion relations are quite
accurate at high energies, detailed studies using derivative dispersion relations \cite{derivative001} would be needed
to quantify the effect of dispersion-relation subtractions on the imaginary part of the eikonal. An analysis using
derivative dispersion relations is in progress.

\section*{Acknowledgments}

We thank F. Dur\~aes, D.A. Fagundes, M. Malheiro, M.J. Menon, A.A. Natale and G. Pancheri for helpful discussions.
This research was partially supported by the Funda\c{c}\~{a}o de Amparo \`{a} Pesquisa do Estado do Rio Grande do Sul
(FAPERGS), Coordena\c{c}\~ao de Aperfei\c{c}oamento de Pessoal de N\'{\i}vel Superior (CAPES) and by Conselho Nacional
de Desenvolvimento Cient\'{\i}fico e Tecnol\'ogico (CNPq). EGSL acknowledges the financial support from RENAFAE.

\newpage

\begin{table*}
\caption{Values of the model parameters from the global fit to the scattering $pp$ and
$\bar{p}p$ data. Results obtained using a monopole form factor in the semihard sector.}
\begin{ruledtabular}
\begin{tabular}{cccc}
 & CTEQ6L & CTEQ6L1 & MSTW \\
\hline
$\nu_{1}$ [GeV] & $1.712\pm0.541$ &  $1.980\pm0.745$ & $1.524\pm0.769$ \\
$\nu_{2}$ [GeV] & (3.376$\pm$1.314)$\times 10^{-2}$ &  (5.151$\pm$1.627)$\times 10^{-2}$ &
(9.536$\pm$8.688)$\times 10^{-3}$ \\
$A'$ [GeV$^{-1}$] & $125.3\pm 14.7$ & $107.3\pm9.0$ & $107.2\pm13.6$ \\
$B'$ [GeV$^{-1}$] & $42.96\pm24.91$ & $28.73\pm14.78$ & $30.54\pm16.20$ \\
$C'$ [GeV$^{-1}$] & $1.982\pm0.682$ & $1.217\pm0.402$ & $1.186\pm0.466$ \\
$\gamma$ & $0.757\pm 0.189$ & $0.698\pm 0.212$ & $0.644\pm0.250$ \\
$\mu^{+}_{soft}$ [GeV] & $0.777\pm 0.176$  & $0.407\pm0.266$ & $0.475\pm0.300$ \\
$D'$ [GeV$^{-1}$] & $23.78\pm 1.97$ & $21.37\pm 2.67$  & $21.92\pm2.83$ \\
\hline
$\chi^{2}/DOF$  & 1.060 & 1.063 & 1.049 \\
\end{tabular}
\end{ruledtabular}
\end{table*}

\begin{table*}
\caption{Values of the model parameters from the global fit to the scattering $pp$ and
$\bar{p}p$ data. Results obtained using a dipole form factor in the semihard sector.}
\begin{ruledtabular}
\begin{tabular}{cccc}
 & CTEQ6L & CTEQ6L1 & MSTW \\
\hline
$\nu_{1}$ [GeV] & $2.355\pm0.620$ & $2.770\pm0.865$ & $2.267\pm0.845$  \\
$\nu_{2}$ [GeV] & (5.110$\pm$4.203)$\times 10^{-2}$ & (7.860$\pm$5.444)$\times 10^{-2}$ & 
(3.106$\pm$2.920)$\times 10^{-2}$ \\
$A'$ [GeV$^{-1}$] & $128.9\pm13.9$ & $108.9\pm8.6$ & $108.5\pm11.5$ \\
$B'$ [GeV$^{-1}$] & $46.73\pm26.13$ & $30.19\pm15.78$ & $31.63\pm16.16$ \\
$C'$ [GeV$^{-1}$] & $2.103\pm0.669$ & $1.260\pm0.437$ & $1.230\pm0.467$ \\
$\gamma$ & $0.780\pm0.170$ & $0.719\pm 0.200$ & $0.660\pm0.227$ \\
$\mu^{+}_{soft}$ [GeV] & $0.821\pm0.150$ & $0.457\pm0.209$ & $0.506\pm0.236$ \\
$D'$ [GeV$^{-1}$] & $23.96\pm1.92$ &  $21.73\pm2.26$ & $22.14\pm2.38$ \\
\hline
$\chi^{2}/DOF$  & 1.064 & 1.062 & 1.047 \\
\end{tabular}
\end{ruledtabular}
\end{table*}

\begin{figure}[!h]
\centering
\hspace*{-0.2cm}\includegraphics[scale=0.70]{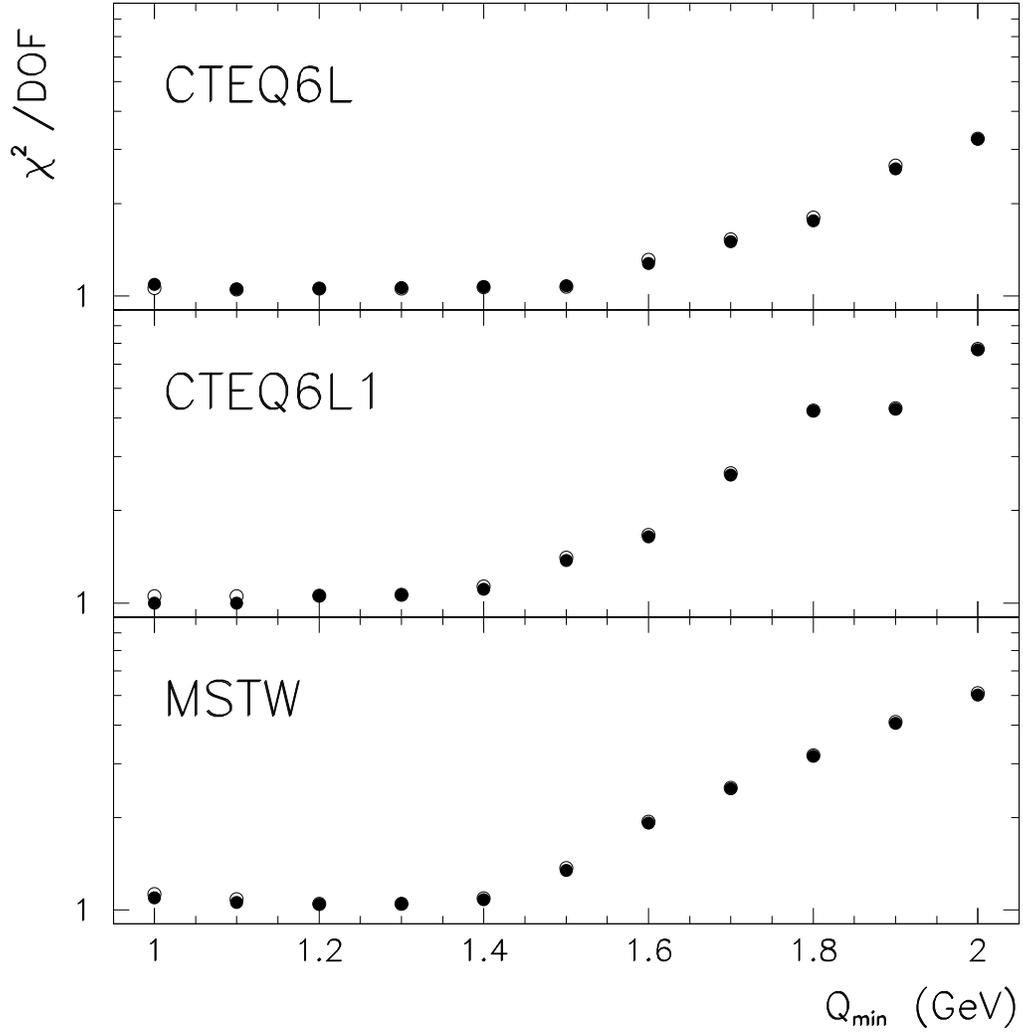}
\label{fig1}
\caption{The $\chi^{2}/DOF$ as a function of the cutoff $Q_{min}$ for the monopole ($\circ$) and the dipole ($\bullet$)
semihard form factor.}
\end{figure}

\begin{table*}
\caption{Predictions for the forward scattering quantities $\sigma_{tot}^{pp,\bar{p}p}$ and $\rho^{pp,\bar{p}p}$ using
different sets of parton distributions.}
\begin{ruledtabular}
\begin{tabular}{cccccc}
 & $\sqrt{s}$ [TeV] & \multicolumn{2}{c}{$\sigma_{tot}$ [mb]} & \multicolumn{2}{c}{$\rho$} \\
 &  &  monopole &  dipole &  monopole & dipole \\
\hline
CTEQ6L & 8.0 &  $100.9^{+8.6}_{-7.3}$ &  $101.0^{+8.6}_{-7.3}$ &  $0.115^{+0.009}_{-0.008}$ & $0.106^{+0.009}_{-0.007}$ \\
 & 13.0 &  $111.5^{+9.7}_{-8.4}$ &  $111.7^{+9.7}_{-8.4}$ &  $0.110^{+0.010}_{-0.008}$ & $0.101^{+0.009}_{-0.008}$ \\
 & 14.0 &  $113.2^{+9.9}_{-8.6}$ &  $113.5^{+9.9}_{-8.6}$ &  $0.110^{+0.010}_{-0.008}$ & $0.100^{+0.009}_{-0.008}$ \\
 & 57.0 &  $152.5^{+15.4}_{-14.7}$ &  $154.1^{+15.6}_{-14.9}$ &  $0.097^{+0.010}_{-0.010}$ & $0.088^{+0.009}_{-0.009}$ \\
 & 95.0 &  $170.3^{+17.2}_{-16.5}$ &  $172.9^{+17.5}_{-16.8}$ &  $0.092^{+0.010}_{-0.010}$ & $0.083^{+0.009}_{-0.009}$ \\
CTEQ6L1 & 8.0 &  $101.1^{+8.6}_{-7.3}$ &  $101.2^{+8.6}_{-7.3}$ &  $0.134^{+0.012}_{-0.009}$ &  $0.124^{+0.011}_{-0.009}$ \\
 & 13.0 &  $112.4^{+9.8}_{-8.5}$ &  $112.9^{+9.8}_{-8.5}$ &  $0.131^{+0.012}_{-0.010}$ &  $0.120^{+0.011}_{-0.009}$ \\
 & 14.0 &  $114.2^{+10.0}_{-8.7}$ &  $114.9^{+10.0}_{-8.7}$ &  $0.130^{+0.012}_{-0.010}$ &  $0.119^{+0.011}_{-0.009}$ \\
 & 57.0 &  $159.3^{+16.1}_{-15.4}$ &  $163.7^{+16.5}_{-15.8}$ &  $0.117^{+0.012}_{-0.012}$ & $0.106^{+0.011}_{-0.011}$ \\
 & 95.0 &  $181.5^{+18.3}_{-17.6}$ &  $188.9^{+19.0}_{-18.4}$ &  $0.112^{+0.012}_{-0.012}$ & $0.101^{+0.011}_{-0.011}$ \\
MSTW & 8.0 &  $101.3^{+8.6}_{-7.3}$ &  $101.3^{+8.7}_{-7.3}$ &  $0.142^{+0.013}_{-0.010}$ &  $0.131^{+0.012}_{-0.009}$ \\
 & 13.0 &  $113.3^{+9.9}_{-8.5}$ &  $113.6^{+9.9}_{-8.5}$ &  $0.139^{+0.012}_{-0.011}$ &  $0.128^{+0.011}_{-0.010}$ \\
 & 14.0 &  $115.4^{+10.1}_{-8.7}$ &  $115.7^{+10.1}_{-8.8}$ &  $0.139^{+0.013}_{-0.011}$ &  $0.128^{+0.012}_{-0.010}$ \\
 & 57.0 &  $162.1^{+16.4}_{-15.6}$ &  $164.7^{+16.6}_{-15.9}$ &  $0.127^{+0.013}_{-0.013}$ & $0.116^{+0.012}_{-0.011}$ \\
 & 95.0 &  $183.0^{+18.5}_{-17.8}$ &  $187.3^{+18.9}_{-18.2}$ &  $0.123^{+0.013}_{-0.013}$ & $0.112^{+0.012}_{-0.012}$ \\
\end{tabular}
\end{ruledtabular}
\end{table*}

\begin{figure}[!h]
\centering
\hspace*{-0.2cm}\includegraphics[scale=0.70]{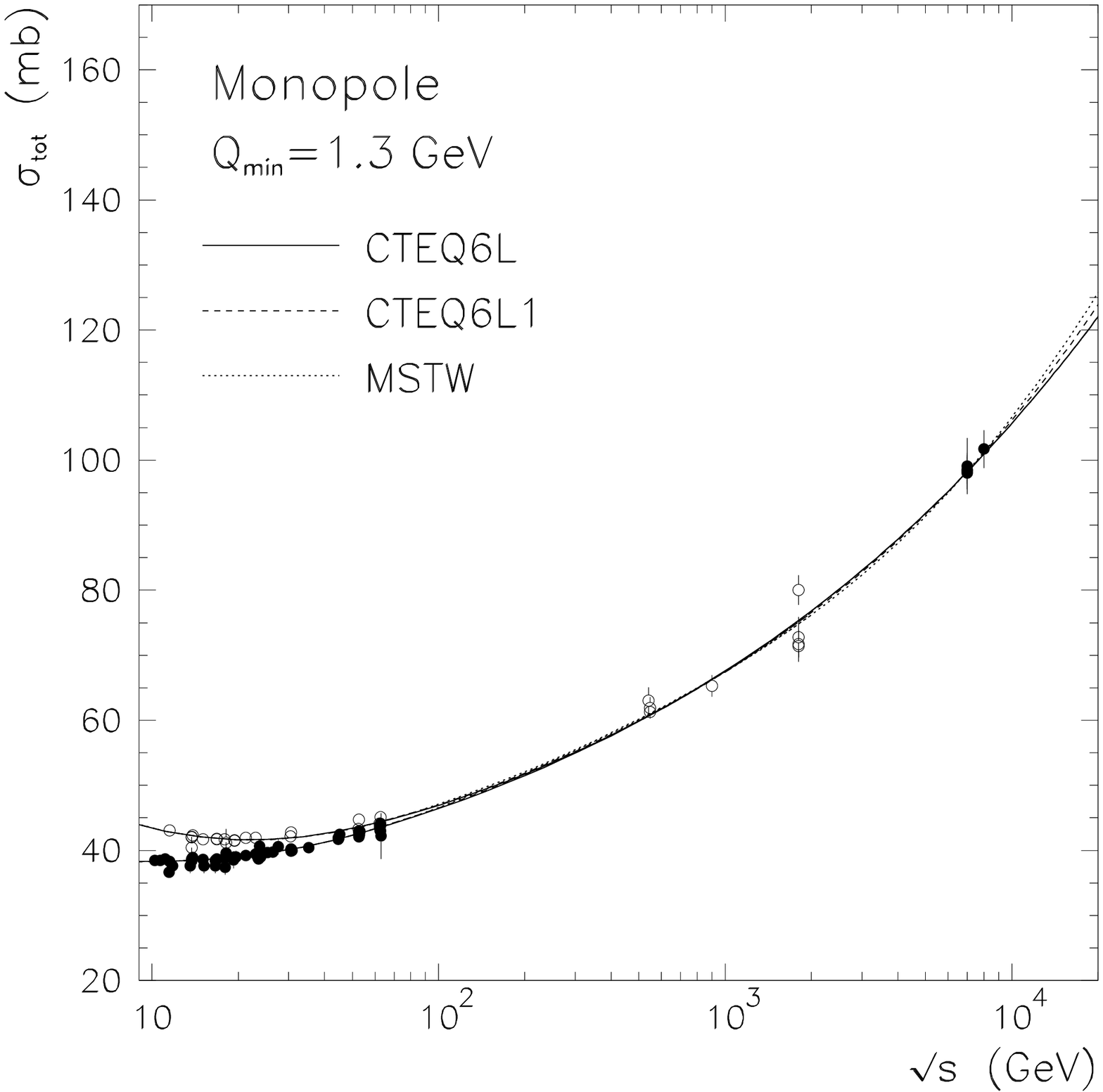}
\label{fig1}
\caption{Total cross section for $pp$ ($\bullet$) and $\bar{p}p$ ($\circ$).}
\end{figure}

\begin{figure}[!h]
\centering
\hspace*{-0.2cm}\includegraphics[scale=0.70]{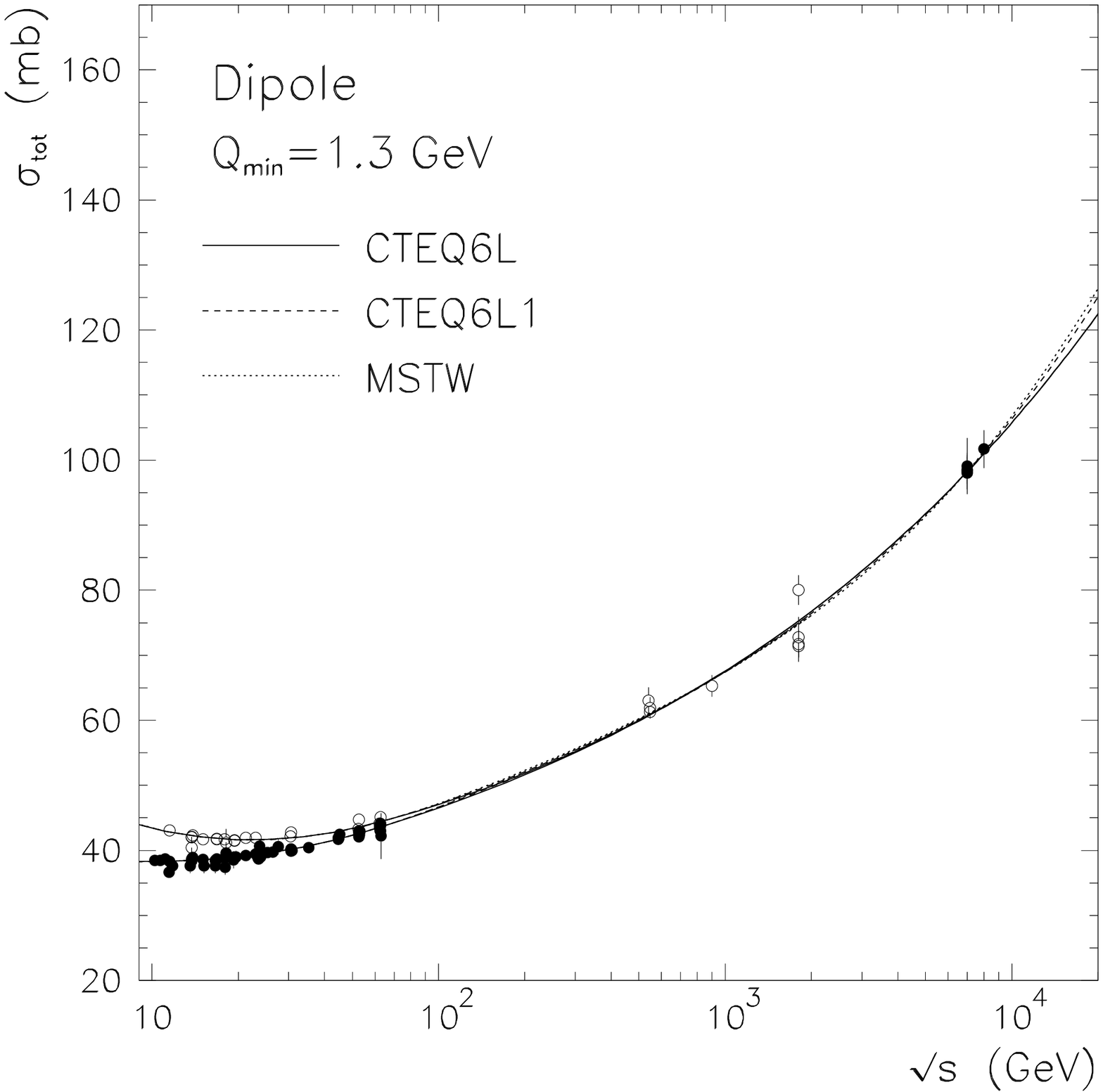}
\label{fig1}
\caption{Total cross section for $pp$ ($\bullet$) and $\bar{p}p$ ($\circ$).}
\end{figure}

\begin{figure}[!h]
\centering
\hspace*{-0.2cm}\includegraphics[scale=0.70]{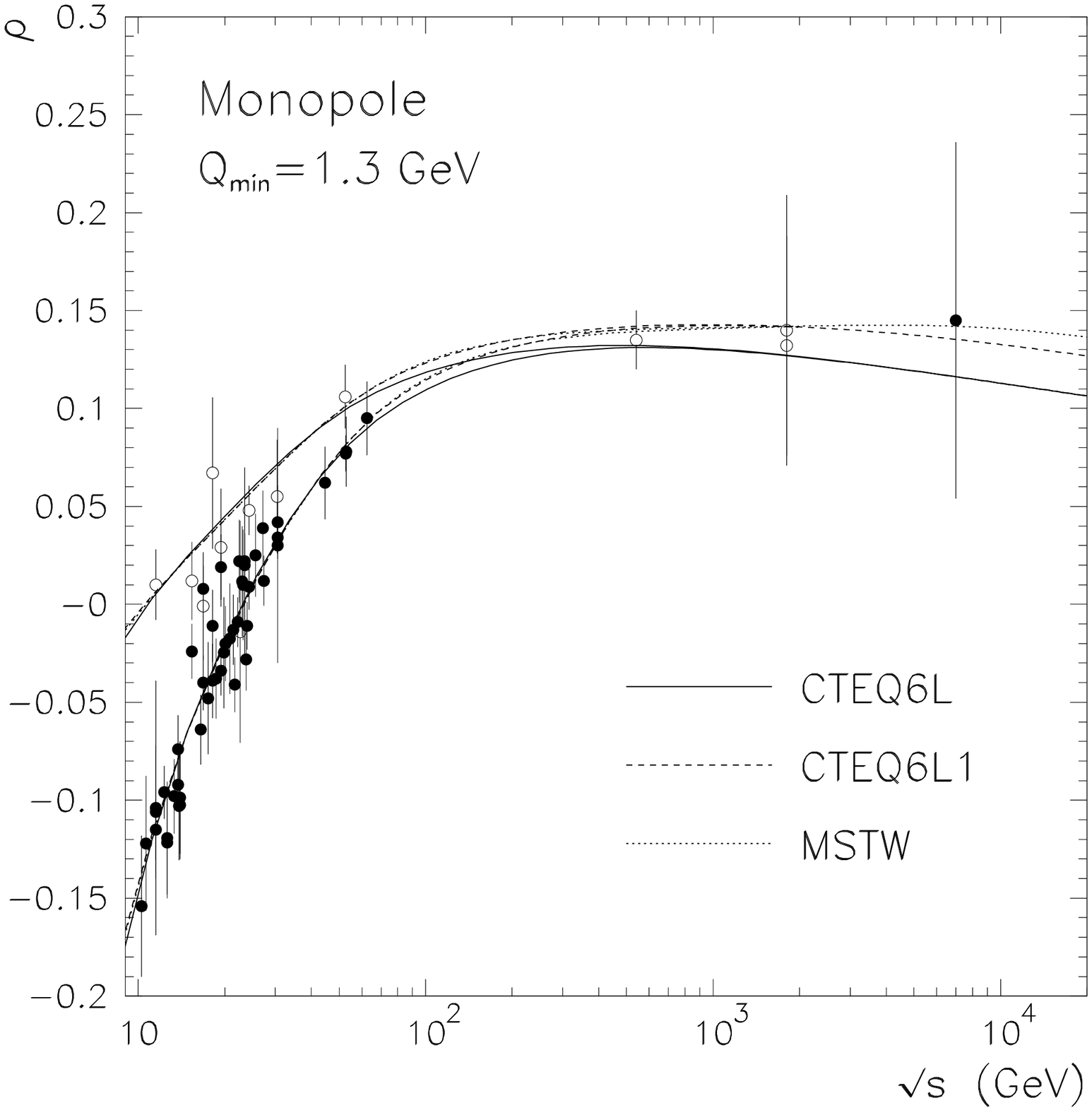}
\label{fig1}
\caption{Ratio of the real to imaginary part of the forward scattering amplitude for $pp$ ($\bullet$) and $\bar{p}p$ ($\circ$).}
\end{figure}

\begin{figure}[!h]
\centering
\hspace*{-0.2cm}\includegraphics[scale=0.70]{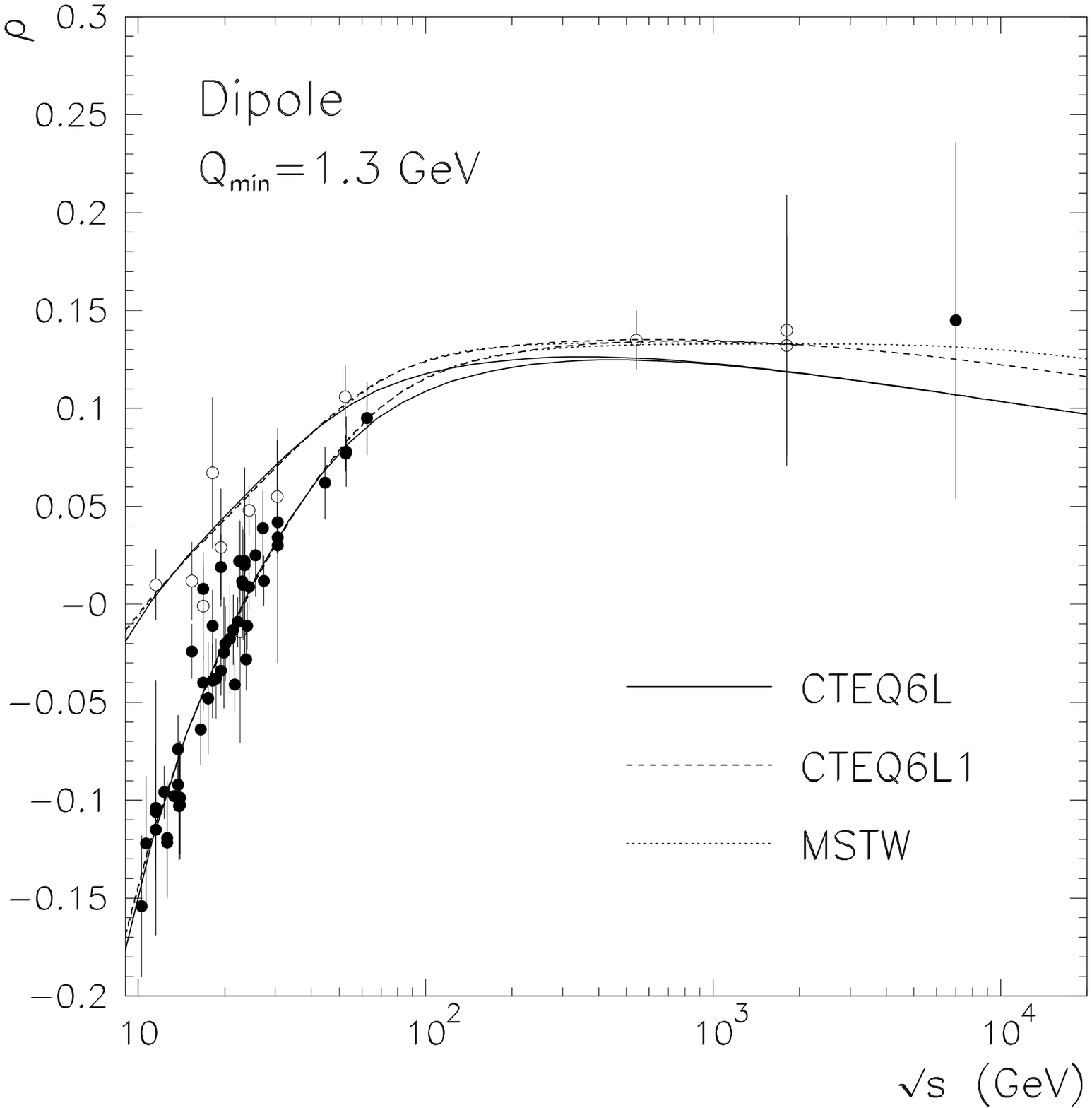}
\label{fig1}
\caption{Ratio of the real to imaginary part of the forward scattering amplitude for $pp$ ($\bullet$) and $\bar{p}p$ ($\circ$).}
\end{figure}

\begin{figure}[!h]
\centering
\hspace*{-0.2cm}\includegraphics[scale=0.70]{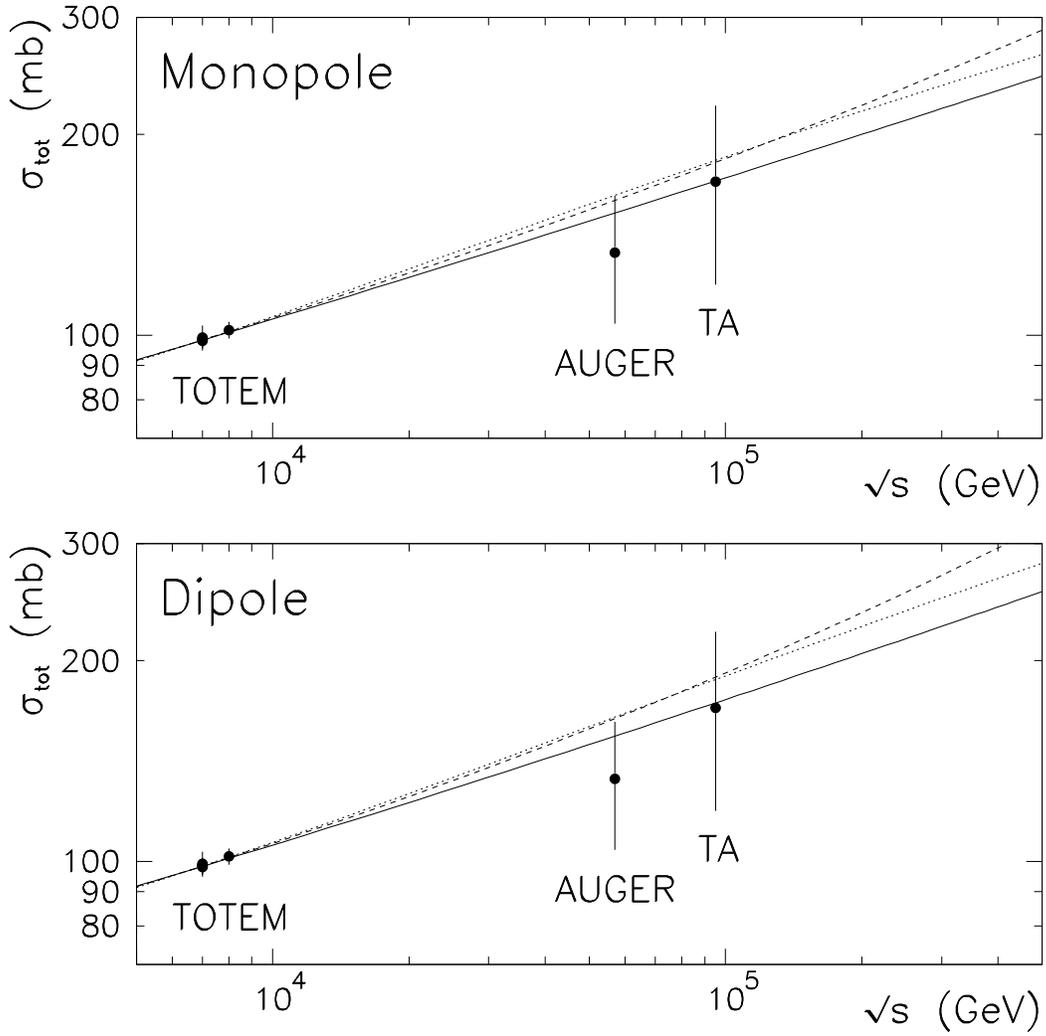}
\label{fig1}
\caption{TOTEM, AUGER and Telescope Array (TA) results compared with theoretical expectations obtained using CTEQ6L
(solid curve), CTEQ6L1 (dashed curve) and MSTW (dotted curve) parton distribution functions.}
\end{figure}

\end{document}